# StegIbiza: New Method for Information Hiding in Club Music

Krzysztof Szczypiorski [1,2]
[1] Cryptomage SA, Wrocław, Poland
[2] Warsaw University of Technology, Warsaw, Poland
e-mail: krzysztof.szczypiorski@cryptomage.io

***Abstract*—In this paper a new method for information hiding in club music is introduced. The method called StegIbiza is based on using the music tempo as a carrier. The tempo is modulated by hidden messages with a 3-value coding scheme, which is an adoption of Morse code for StegIbiza. The evaluation of the system was performed for several music samples (with and without StegIbiza enabled) on a selected group of testers who had a music background. Finally, for the worst case scenario, none of them could identify any differences in the audio with a 1% margin of changed tempo.**

***Keywords: information hiding, audio steganography, musical steganography, club music, StegIbiza***

*"Be free with your tempo, be free, be free.*
*Surrender your ego be free, be free to yourself."*
Queen – *"Innuendo"*

## I. INTRODUCTION

Steganography seems to be a very attractive subject area for sharing information via the Internet around the globe without (if possible) any trace. Recently, major attention has been paid to constructing image [1] and network [2] steganography methods. Lately, less effort has been applied to audio steganography [3], so this work revisited this attractive area for research right now. This paper will focus on musical steganography, which is a part of audio steganography depending naturally on a melody itself and the musical instruments, including the voice as an instrument. Especially, when applying information hiding to music streaming services, this seems to be very promising as there is an almost infinite source of traffic to hide in.

In this paper, a brand new method of information hiding in club music, called StegIbiza (Steganographic Ibiza), is introduced. Ibiza is the third largest of the Balearic Islands (Spain) in the western Mediterranean Sea, near the eastern coast of the Iberian Peninsula. Ibiza is also one of the top holiday destinations in Europe that is best known for its nightclub based nightlife. The city has achieved worldwide fame as a cultural center for club music, including house and trance.

The proposed method of StegIbiza (Figure 1) is based on the music tempo measured in beats per minute (bpm). Tempo means "time" in Italian and is one of the major parameters of mixed music by DJs (disc jockeys) in compositions called mashups, which are created by blending two or more pre-recorded songs. In blending music from two or more decks (i.e., CD/MP3 players), the tempo must be synchronized to hide the blending process.

In StegIbiza the tempo of the music is modulated by hidden messages with a 3-value coding scheme, which is an adoption of Morse code. It allows for changes in the tempo that are very delicate, like +/- 1-2 bpm from the original tempo, which was taken as a reference to be undetectable by human hearing. For club music a typical reference tempo is 100-150 bpm and usually constant for a whole mushup.

To prove that StegIbiza is inaudible to humans, several music samples (with and without StegIbiza enabled) were prepared using a professional digital audio workstation (DAW) and then an evaluation was performed on a selected group of testers also who had a music background.

This work was inspired partially by electrocardiography (ECG) steganography ([4], [5]), which was called in an IEEE Spectrum article, "hiding data in a heartbeat" [6]. These methods involve electrical signals from a heart being used to carry hidden data. The second source of inspiration was a song by Enrique Iglesias featuring Nicole Scherzinger – "*Heartbeat*" (with lyrics: *"I can feel your heartbeat. Running through me"*).

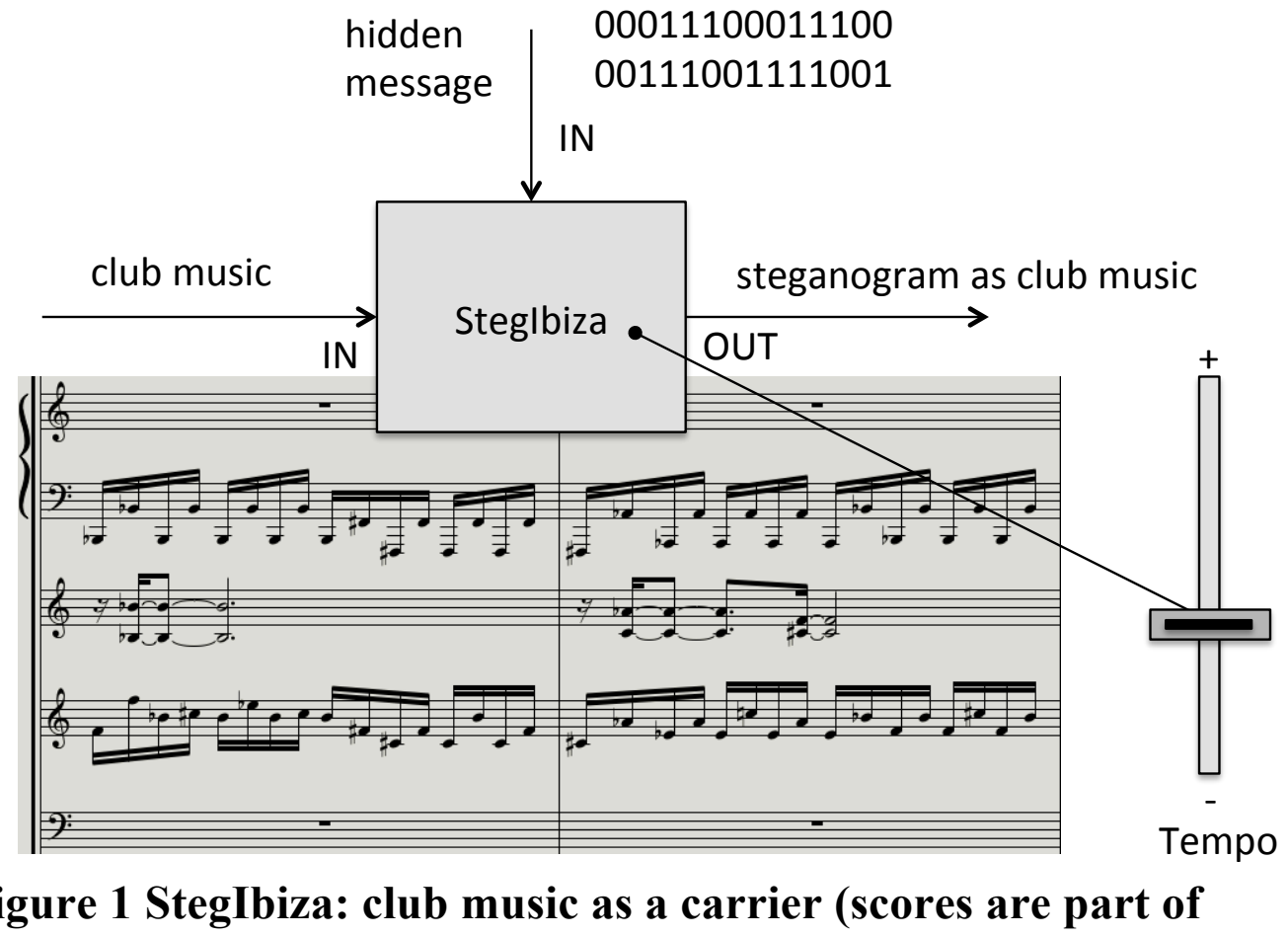

**Figure 1 StegIbiza: club music as a carrier (scores are part of Snap! – "*Rhythm Is a Dancer*").**

The paper is structured as follows: Section II briefly presents the state of the art in audio steganography, including musical steganography. Section III contains a presentation of the idea of the StegIbiza method and two typical scenarios for preparation of steganograms: one with a DAW and one with manual changing of the tempo by a DJ. In Section IV the work describes the testing environment as a proof of concept and shows the results of the evaluation for these scenarios. Section VI includes a discussion on the possibilities of detection of the proposed system. Finally, Section VII concludes our efforts and contains future work.

## II. State of the Art in Audio Steganography

Considering that at the begin of the XVI century Johannes Trithemius in "*Polygraphiae*" [7] presented a method based on lyrics in "*Ave Maria virgo serena*" (dated ca. 1485) and corresponding secret tables, we can assume that musical steganography for exchanging hidden information has been known and used for at least 500 years.

In 1665, Gaspar Schott in "*Schola steganographica*" [8] presented how to use music scores for information hiding – each note is mapped into a letter. Based on this idea, many systems were introduced, sometimes known as musical cryptograms [9]. The most recognized was a BACH motif (B♭, A, C, B♮), which was used by Johann Sebastian Bach himself and by many other composers. These methods could be used as a naïve encryption system or just as a watermarking method sometimes to honor other composers. Some examples are as follows:

- B♭, A, B, E♭ (= B, r, A, H, m, Es) for Johannes Brahms.
- B, E, B, A or B, A, B, E for Béla Bartók.
- C, A, G, E for John Cage.
- F, E♭, C, B (= F, S, C, H) for Franz Schubert.
- D, E♭, C, B (= D, S, C, H) for Dmitri Shostakovich.

From the perspective of audio steganography, the major innovations were recording of the voice and music and then broadcasting them. The analogue era introduced the possibility of adding some noise and other signals to the transmission [10].

After the transformation of the audio world from analog to digital, five major methods [11] of audio steganography were introduced:

- Least Significant Bit coding [12]: a binary sequence of each sample of a digitized audio file is replaced with a binary equivalent of a hidden message.
- Parity coding [13]: instead of dividing a part of a signal into individual samples, the parity coding method breaks a signal down into separate regions and encodes each bit from the hidden message in a sample region's parity bit.
- Phase coding [14]: the phase components of sound are not as perceptible to the human ear as noise, so in this method the phase of the subsequent segments is then adjusted to preserve the relative phase between the segments.
- Spread spectrum [14]: this is a method of spreading out the encoded data across the available frequencies as much as possible.
- Echo hiding [15]: a hidden message can be embedded in audio data by introducing an echo to the original signal by varying three parameters of the echo: initial amplitude, decay rate, and offset.

There have been some efforts on the connection between beats and watermarking: beat detection to combat desynchronization and watermark estimation [16] and constructing beat-based watermarks [17].

## III. Idea of StegIbiza

The idea of StegIbiza is based on the modulation of the music tempo by hidden messages. In the initial implementation the hidden messages were encoded with an adoption of Morse code (Table I). Morse code was chosen for two reasons: it is very popular in many communities and by definition is very easy to use by humans.

TABLE I. Morse code adopted for StegIbiza.

| Character | Code | Character | Code |
|---|---|---|---|
| a | +- | 1 | +---- |
| b | -+++ | 2 | ++--- |
| c | -+-+ | 3 | +++-- |
| d | -++ | 4 | ++++- |
| e | + | 5 | +++++ |
| f | ++-+ | 6 | -++++ |
| g | --+ | 7 | --+++ |
| h | ++++ | 8 | ---++ |
| i | ++ | 9 | ----+ |
| j | +--- | , | --++-- |
| k | -+- | . | +-+-+- |
| l | +-++ | : | ---+++ |
| m | -- | ; | -+-+-+ |
| n | -+ | ! | -+-+-- |
| o | --- | ? | ++--++ |
| p | +--+ | ' | +----+ |
| q | --+- | - | -++++- |
| r | +-+ | _ | ++--+- |
| s | +++ | / | -++-+ |
| t | - | ( | -+--+ |
| u | ++- | ) | -+--+- |
| v | +++- | " | +-++-+ |
| w | +-- | = | -+++- |
| x | -++- | + | +-+-+ |
| y | -+-- | & | +-+++ |
| z | --++ | @ | +--+-+ |
| 0 | ----- | $ | +++-++- |

In the adoption of Morse code for StegIbiza a "dot" is replaced with a "+" (a "plus" symbol) and a "dash" is left as it was ("–", a "minus" symbol). A "plus" means a change in the tempo of $+\Delta$ bpm for a period of $\Phi$ beats, and the "minus" means a change of $-\Delta$ bpm for the same period of $\Phi$ beats. The original tempo was taken as a reference and corresponds to a silence ("0", a "zero" symbol), so it creates a 3-value variable-length code with a comma. Other important information about the adopted Morse code are as follows:

- A "plus" is one unit long, which corresponds to $\Phi$ beats.
- A "minus" is one unit long ($\Phi$).
- An intra-character gap (between the "pluses" and "minuses" within a character) is not present.
- The space between two letters is one unit long ($\Phi$).
- The space between two words is equal to two units ($2\Phi$).

An example of the use of the StegIbiza method for Φ = 1 beat and Δ =1 bpm with a hidden message containing "a" (+-) is presented in Figure 2.

Now, it is possible to formulate the following research questions: what should be (Φ, Δ) to give a change in the tempo that is inaudible to humans? What should be the characteristics of the signal edge?

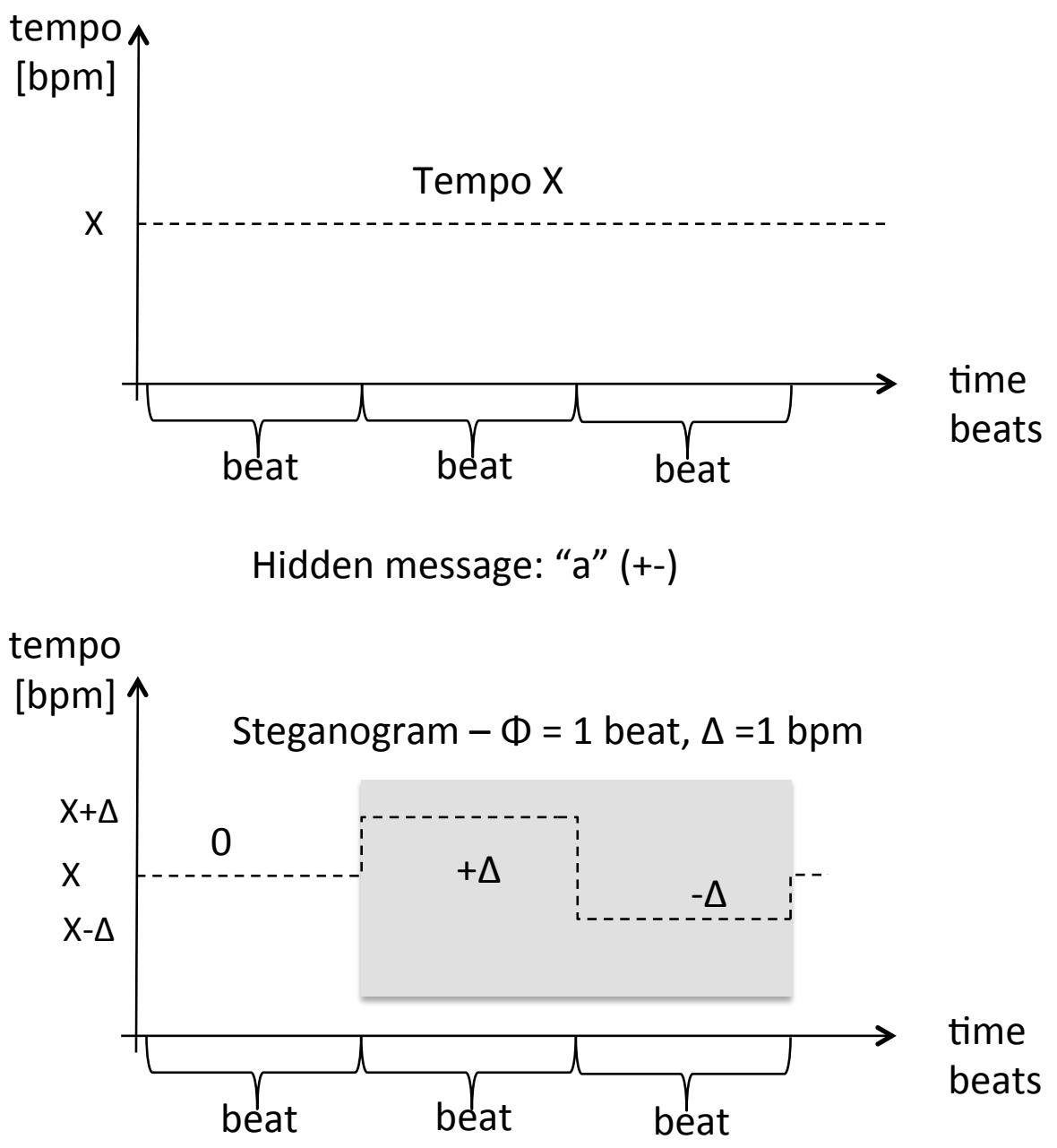

**Figure 2 Explanation of StegIbiza method (Φ = 1 beat, Δ =1 bpm) – hidden message: "a" (two symbols: "+" and "-").**

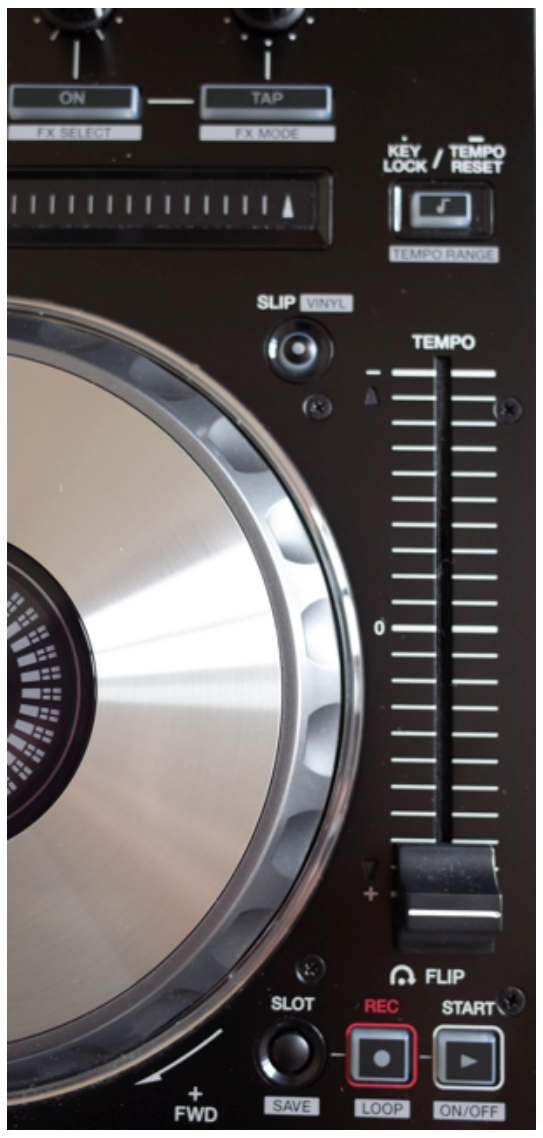

**Figure 3 Tempo slider (on the right) in a DJ controller (Pioneer DDJ-SX2).**

Now consider two scenarios of preparing steganograms (off-line and on-line) that would help to initially answer these questions. In both scenarios a streaming music service could transfer steganograms around the world (!). It could be on-demand or a classical Internet service available on many existing broadcasting platforms.

In an off-line scenario, make mashups and, based on a DAW, tune the desired tempo with no limits over the shape of the signal edge (could be rectangular) and (Φ, Δ).

In an on-line scenario, use a DJ controller (a device used to help DJs mix music) with a tempo slider (Figure 3). Typically there are also no limits to the level of the tempo (Δ), but the signal shape as well as the minimal duration of Φ depends on the manual skills of the DJ. In this, a DJ will be a contemporary telegrapher with a DJ controller as the telegraph.

## IV. PROOF OF CONCEPT AND EVALUATION

For the proof of concept this work used a professional DAW: Logic X Pro by Apple. The latest version (10.2.4) was run with all available extensions on an Apple OS X El Capitan (10.11.6) machine with a 2.8 GHz Intel Core i5 processor, 8 GB 1600 MHz DDR3 of RAM, and an Intel Iris 1536 MB graphic card.

Based on MIDI (Musical Instrument Digital Interface) files this work prepared five covers of popular songs with a constant tempo:

- *"Lily was here"* by David A. Stewart and Candy Dulfer (131 bpm).
- *"Miracle"* by Queen (92 bpm).
- *"Rhythm is a dancer"* by Snap! (130 bpm).
- *"So what"* by Miles Davis (120 bpm).
- "*You were the heart's beat"* (in Polish: *"Byłaś serca biciem"*) by Andrzej Zaucha (100 bpm).

All original covers were prepared without any vocal parts and arranged in techno, hip-hop, or trance styles with the instruments available in Logic X Pro.

The chosen hidden message was "*steganography is a dancer!*", which when encoded into a Morse code looks like:

```
... - . --. .- -. --- --. .-. .- .--. .... -.--
.. ...       .-          -.. .- -. -.-. . .-. -.-.--
```

and then adopted to the StegIbiza (Figure 4) containing 88 symbols:

```
+++0-0+0--+0+-0-+0---0--+0+-+0+-0+--+0++++0-+--
00++0+++00+-00-++0+-0-+0-+-+0+0+-+0-+-+--
```

The evaluation of the first scenario was for Δ∈{0.5, 1, 1.5, 2, 2.5, 3} bpm with a rectangular shape and constant Φ = 1 beat. This work embedded a hidden massage to random parts of each song (twice) so there were 60 tracks for evaluation. All of them were at the highest quality and uncompressed.

The group of testers (n=20) consisted of 11 women and 9 men. They were aged 25-45 and did not have any hearing impairments. Seven of them had a professional music background (graduated from musical school) and three of them worked as professional musicians. For any Δ that gives a 1% difference in tempo (counted as Δ/X, where X is the original tempo, Figure 2) – none of them could identify a difference in the audio. For above 1% but below 2%, only the professional musicians could identify the differences. Above 2%, around 50% (9) of the participants found differences. For 3% and above, all of the participants detected the StegIbiza. In the experiment, all the participants used Hi-Fi professional headphones (they could choose from three types from two vendors: two Sennheiser devices and one from Bose). After passing a blind test with the StegIbiza enabled, the location of the hidden message was shown to them. Even in this case they did not claim to be able to hear any differences. The selected tests were repeated with compressed audio files (MPEG-4 Audio Layer 128 kBit/s and 256 kBit/s, and there was no change in the results).

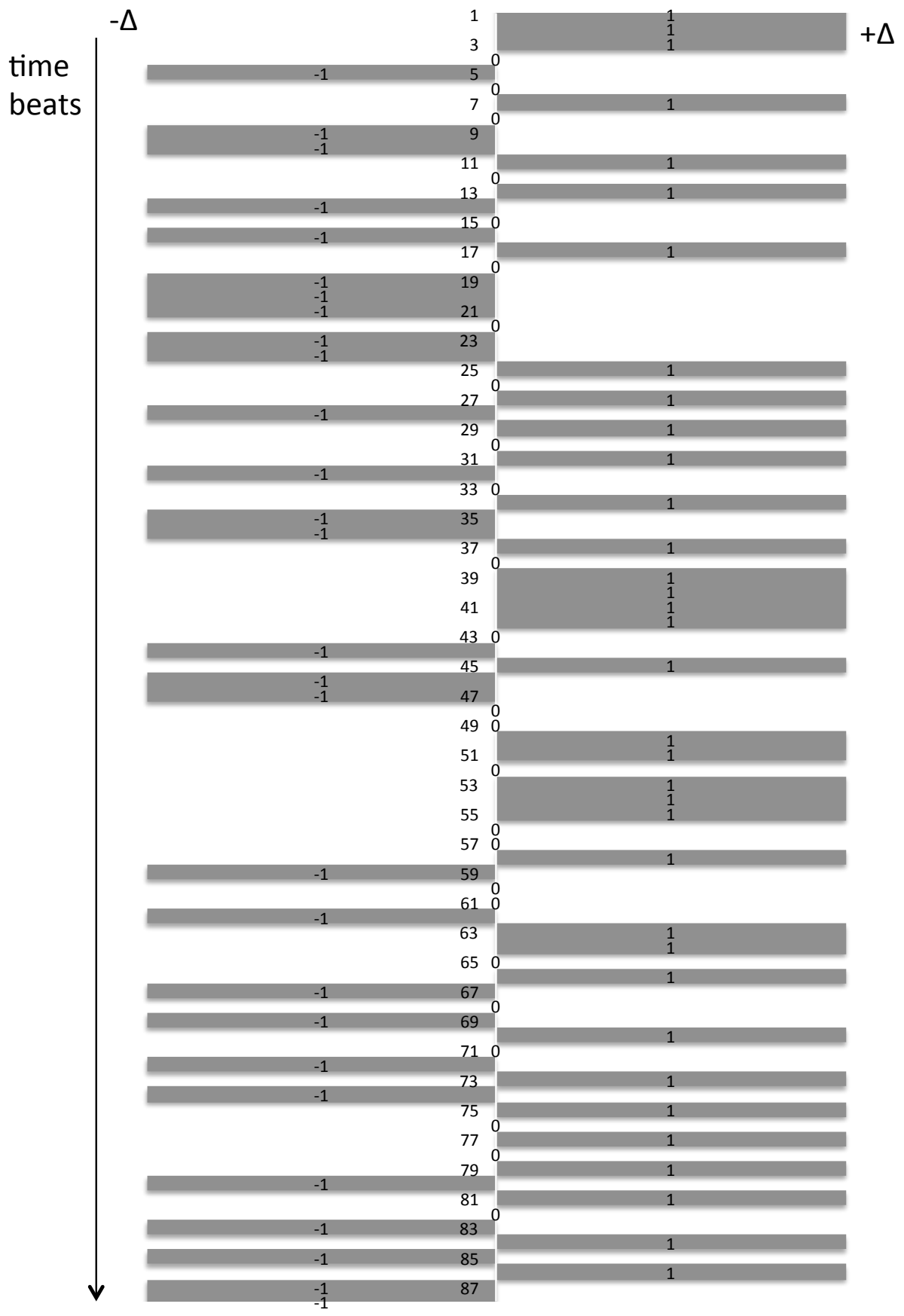

**Figure 4 StegIbiza method (Φ = 1 beat, Δ = 1 bpm) – hidden message: "steganography is a dancer!".**

The evaluation for the second scenario was made during a summer open air party. The group of testers (n=134) consisted of 69 women and 65 men. They were aged 18-43 with no information about any hearing impairments. For this evaluation the work used only a tempo slider (from -8% to +8%, i.e., ca. -5 < Δ < +5) in a DJ controller (Pioneer DDJ-SX2 with Serato software) for regular MP3 files with top club music remixes. It was tested for Φ=3 and more with linear characteristics. The environment (echo, DJ effects) had a tremendous impact on the audience and the following results were obtained: below 2% nobody identified any differences in tempo. Above 2% but below 3%, around ¼ of the participants identified differences. Above 3% but below 4%, around ½ of the participants identified differences. At this level the experiment was stopped, because the rest of the party did not care about the music.

## V. Discussion

This paper is a report on work in progress rather than a publication of final results, so there will now be a discussion in this section about some issues concerning assumptions and the security of the proposed system.

For the first scenario, Φ = 1 as a logical starting point – size of one beat - as smaller values are not interesting for evaluation. One beat is a solid part of a rhythm and better results than with Φ < 1 were not expected. For the second scenario, Φ = 3 as that was the limit of the DJ's skills and it could not be made quicker. However, a larger Φ means less speed, but it should be more inaudible.

The "Δ" parameter is mainly linked with the environment for the reception of the club music. For evaluation of the first scenario, professional headphones and high quality source signals were used, so for a regular audience the results should be better (more than 1% margin of changed tempo).

This work presents two scenarios for the preparation of the steganograms, but one can imagine developing separate software to implement the StegIbiza method. It could be done in two ways: by usage of a DAW's API (Application Programming Interface), for example, through plug-in interfaces, or by writing specialized software (with some existing audio libraries like Aubio [18]).

StegIbiza could be detected automatically by algorithms for beat detection as a part of the music information retrieval. There are special annual contest for this type of algorithm, for example, MIREX (Music Information Retrieval Evaluation eXchange – [19]). Many of these algorithms are based on a Fourier transformation to find peaks in the spectrum of the signal and then to perform a deeper analysis of the sonograms that were made ([20], [21]).

In [22], I presented with my colleagues a classification of steganography methods with three levels of undetectability, named: "good", "bad", and "ugly". According to this categorization (which was formally proposed for network steganography, but that could be extended to all other methods with data in motion), StegIbiza seems to be an "ugly" method, as the observer is able to detect the hidden communication anywhere in the network, even at the steganographic receiver of the hidden data. However, most steganographic algorithms are “ugly”, so from the perspective of detection and protection against these kind of threats, it seems to have a positive rank.

## VI. Conclusions and Future Work

This article presented the StegIbiza method for information hiding in club music based on the tempo as a carrier. It showed how to embed hidden messages with a special 3-value adopted from Morse code, and then it presented two practical scenarios for the preparation of steganograms – one DAW based and one DJ controller based. Both scenarios were evaluated with a reference group, and it was found that an audience did not find any differences with a 1% margin of changed tempo.

Planned future steps are as follows:

- A dedicated tool to change the tempo in any music stream when using the StegIbiza method.
- A plugin for IDS/IPS software to detect all tempo based threats that are similar to the proposed system.

In addition, other coding schemes, than that proposed in this paper of a 3-value adoption of Morse code, will be considered.

## Acknowledgments

I would like to thank all the people involved in the experiments presented in this paper. I appreciate your time and entanglement in this kind of research and hope that it was a good opportunity for joy and a meeting. I promise I will improve my DJing skills soon.

This is an independent publication and has not been authorized, sponsored, or otherwise approved by Apple, Bose, Intel, Pioneer, Sennheiser, and Serato. I also have no personal interest to promote Ibiza as a tourist destination.

## References

[1] J. Fridrich, “Steganography in Digital Media: Principles, Algorithms, and Applications”, Cambridge University Press; 1 edition, December 2009

[2] W. Mazurczyk, S. Wendzel, S. Zander, A. Houmansadr, K. Szczypiorski, “Information Hiding in Communication Networks: Fundamentals, Mechanisms, Applications, and Countermeasures”, Wiley-IEEE Press; 1 edition, February 2016

[3] N. Cvejic, “Algorithms for audio watermarking and steganography”, Doctoral Dissertation, Acta Universitatis Ouluensis, C Technica, University of Oulu, 2006

[4] A. Ibaida, I. Khalil, "Wavelet-Based ECG Steganography for Protecting Patient Confidential Information in Point-of-Care Systems", in IEEE Transactions on Biomedical Engineering, vol. 60, no. 12, pp. 3322-3330, December 2013

[5] S. Jero, P. Ramu, S. Ramakrishnan, “Discrete wavelet transform and singular value decomposition based ECG steganography for secured patient information transmission”, Journal of Medical Systems, 38:132, October, 2014

[6] L. Newman, “Hiding Data in a Heartbeat”, IEEE Spectrum, 10 Oct 2013, http://spectrum.ieee.org/tech-talk/biomedical/diagnostics/hiding-data-in-a-heartbeat

[7] J. Trithemius, “Polygraphiae”, 1518

[8] G. Schott, “Schola steganographica”, 1665

[9] E. Sams, “Musical cryptography”, Cryptologia, vol. 3, no. 4, (1979), pp. 193-201

[10] D. Khan, “The Codebreakers: The Comprehensive History of Secret Communication from Ancient Times to the Internet”, Scribner, December, 1996

[11] N. Johnson, S. Katzenbeisser. "A survey of steganographic techniques." Information hiding. Norwood, MA: Artech House, 2000.

[12] S. Dumitrescu, X. Wu, Z. Wang, “Detection of LSB steganography via sample pair analysis”, IEEE Transactions on Signal Processing, vol. 51, no. 7, pp. 1995-2007, July 2003

[13] Y. Kim, Z. Duric, D. Richards, “Limited distortion in LSB steganography”, Proc. SPIE 6072, Security, Steganography, and Watermarking of Multimedia Contents VIII, 60720N, February 16, 2006

[14] W. Bender, D. Gruhl, N. Morimoto, A. Lu, "Techniques for data hiding," in IBM Systems Journal, vol. 35, no. 3.4, pp. 313-336, 1996

[15] D. Gruhl, A. Lu, W. Bender, "Echo hiding", Proc. of First International Workshop Cambridge, U.K., May 30 – June 1, 1996, Volume 1174 of the series Lecture Notes in Computer Science, pp. 295-315, 1996

[16] D. Kirovski, H. Attias, “Audio Watermark Robustness to Desynchronization via Beat Detection”, 5th International Workshop, IH 2002 Noordwijkerhout, The Netherlands, October 7-9, 2002, Volume 2578 of the series Lecture Notes in Computer Science, pp. 160-176, 2002

[17] W. Li, X. Zhang, Z. Wang, “Music content authentication based on beat segmentation and fuzzy classification”, EURASIP Journal on Audio, Speech, and Music Processing (2013) 2013: 11

[18] Aubio library, http://aubio.org, (last visited: 6th of August 2016)

[19] MIREX, http://www.music-ir.org/mirex/wiki/MIREX_HOME, (last visited: 6th of August 2016)

[20] G. Tzanetakis, G. Essl, P. Cook, "Audio analysis using the discrete wavelet transform", Proc. Conf. in Acoustics and Music Theory Applications. 2001

[21] M. Goto, Y. Muraoka, "Real-time beat tracking for drumless audio signals: Chord change detection for musical decisions." Speech Communication 27.3 (1999), pp. 311-335

[22] K. Szczypiorski, A. Janicki, S. Wendzel: “The Good, The Bad And The Ugly”: Evaluation of Wi-Fi Steganography”, Journal of Communications (JCM), Vol. 10(10), pp. 747-752, 2015.